\documentclass{pasj00}
\begin{document}
\SetRunningHead{J. Nakashima, S. Deguchi, and N. Kuno}{Bipolar Nebula IRAS 19312+1950}
\Received{2003/04/01}%{yyyy/mm/dd}
\Accepted{2003/12/24}%{yyyy/mm/dd}

\title{Study of the Bipolar Nebula IRAS 19312+1950. I. \\ Mapping Observations}

%%% begin:list of authors
\author{Jun-ichi \textsc{Nakashima}%
\thanks{ Present adress:  Department of Astronomy, University of Illinois 
at Urbana-Champaign, 1002 W Green St., Urbana, Illinois 61801-3080, U. S. A.}
}
\affil{Department of Astronomical Sciences, The Graduate University for Advanced Studies, \\
Nobeyama Radio Observatory, Minamimaki, Minamisaku, Nagano 384-1305}
\email{junichi@astro.uiuc.edu}

\author{Shuji \textsc{Deguchi} and  Nario \textsc{Kuno}}
\affil{Nobeyama Radio Observatory, National Astronomical Observatory,\\
Minamimaki, Minamisaku, Nagano 384-1305}
%\email{deguchi@nro.nao.ac.jp}
\author{PASJ 56 No. 1 (2004/02/25 issue) in press}
%\affil{C-Address of Institute}\email{ccccc@xxx.xxx.xx.xx}
%%% end:list of authors

%%% Please use the following style in case that sorting by 
%%% affilation is impossible. 
%
% \author{%
%   D-Firstname \textsc{D-Familyname}\altaffilmark{1}
%   E-Firstname \textsc{E-Familyname}\altaffilmark{1,2}
%   and
%   F-Firstname \textsc{F-Familyname}\altaffilmark{2}}
% \altaffiltext{1}{Address of Institute}
% \email{ddddd@xxx.xxx.xx.xx}
% \email{eeeee@xxx.xxx.xx.xx}
% \altaffiltext{2}{Address of Institute}

%% `\KeyWords{}' always has to be placed before `\maketitle'.
 %Do NOT move this preamble from here!

\maketitle
%\setcounter{page}{0}
%%%%% Abstract %%%%%
\begin{abstract}
 IRAS 19312+1950  is an SiO maser source that exhibits a prominent bipolar nebulosity. Mapping observations of this object were made in the CO $J=1$--$0$, $^{13}$CO $J=1$--$0$, C$^{18}$O $J=1$--$0$, CS $J=2$--$1$, and HCN $J=1$--$0$ lines and in the 150 GHz continuum band.  Near-infrared imaging observations were also made in the $J$, $H$, and $K$-bands. The line profiles of the $^{12}$CO and HCN spectra consist of a weak broad component with a line width of about 50 km s$^{-1}$ and a strong narrow component of the width of about 3 km s$^{-1}$. The profiles of the $^{13}$CO, C$^{18}$O, and CS lines have only the narrow component.  Both of the components have an intensity peak at the IRAS position. The narrow component was clearly resolved with a $15''$ telescope beam. The spectral energy distribution of this object exhibits a doubly peaked profile between 1 and 25 $\mu$m. The 150 GHz continuum flux density was found to be 0.07 Jy, which is consistent with the flux density predicted by the expanding envelope model with a mass loss rate of $\sim 10^{-4}$ $M_{\odot}$ y$^{-1}$ at a distance of 2.5 kpc. We argue that the broad component originates from the expanding envelope of this object, and that the hot dust cloud, which is the source of the narrow component, is also physically associated with this object. Though the present observations  do not preclude the possibility of a young stellar object,  we argue that it is less plausible. We conclude that IRAS 19312+1950 is an AGB/post-AGB star that is evolved from a massive progenitor.
\end{abstract}

%%%%% Introduction %%%%%
%\newpage
\section{Introduction}
Protoplanetary nebulae (PPNe) are transient objects between the asymptotic-giant-branch (AGB) 
and planetary-nebula phases of stellar evolution. They are key objects for understanding the formation 
of planetary nebulae. Though OH and H$_2$O masers have been extensively searched for 
in late-AGB stars and PPNe (e.g., \cite{lik89}; \cite{hu94}),  relatively little work has been made 
in  SiO masers. Observations of OH/IR stars and objects with cold envelopes  were made 
in SiO masers by \citet{jew91}, Nyman et al. (1986, 1993),  \citet{gom90},  and  \citet{nak03}.  
It was found in these works that the detection rate of SiO and H$_2$O masers drops with an increase 
of the IRAS [25--12] color, which is a temperature/evolutionary-stage indicator of the dust envelopes 
of late-type stars.

Though H$_2$O and OH masers are often found in PPNe, SiO masers are rarely detected in these objects. The SiO masers are emitted in circumstellar shells very close to the central star ($\sim 10^{14}$ cm), and the OH masers are further away ($\sim 10^{16}$ cm). Therefore, when the mass loss terminates, the SiO masers disappear first, whereas the OH masers remain active during the protoplanetary and early planetary-nebula stages (\cite{lew89}). Thus, only a few PPNe with SiO masers have been found in the past; for example, OH 231.8+1.4 (e.g., \cite{bar84}; \cite{nym98}). 
\citet{nak03} newly found SiO masers in such cold objects. In particular, IRAS 19312+1950 has a prominent bipolar nebulosity with a large angular size of more than 40$''$ on the 2MASS near-infrared images. 
% This is the second example of the bipolar nebulae with SiO masers after OH 231.8+4.2. 

The object, IRAS 19312+1950, is unusual in several respects.  Firstly, it appears to be oxygen-rich (O-rich) according to the detections of SiO, H$_2$O, and OH masers (\cite{nak00}) and the thermal line of SO (\cite{deg03}). But, in addition, the thermal line of H$^{13}$CN was also detected toward this object (\cite{nak00}).  This molecule is normally considered to be an indicator of carbon-rich (C-rich) circumstellar chemistry under the equilibrium condition, though several exceptions have been found at the non-equilibrium (\cite{deg85}; \cite{deg86}). Secondly, a doubly peaked profile was observed in the spectrum of the SiO maser in the $J=$1--0, $v=$2 transition, and SiO maser emission exhibits a strong time variation (\cite{deg03}). The spectral features of OH masers are also quite unusual, exhibiting multiple spiky peaks around the systemic velocity (B. M. Lewis 2000,  private communication).

This object has a cool dusty envelope; the IRAS color, $\log (F_{25}/F_{12})=0.47$, indicates the presence of cool dust with a temperature of about 200 K around the central star. This IRAS color is at the boundary between AGB and post-AGB phases in the two-color diagram (\cite{nym98}). Though we postulated that this is a protoplanetary nebula, it is reasonable to think if this is a young stellar object with SiO masers in a molecular cloud,  
like IRc 2 in Orion.  In fact, \citet{nak00} discussed  a double-peak SiO maser profile of this object,
which is relatively rare in AGB stars, but is well known in Ori IRc 2. However, based on the various observational 
facts that have been collected since the discovery, we believe that this is an evolved object. A discussion on this point will be given in section 3 of this paper and in a forthcoming paper.

Because IRAS 19312+1950 has both prominent and unparalleled characteristics as a protoplanetary nebula, we decided to observe this object more carefully. This object may provides an important clue to solve a puzzle of mass ejection mechanism in a class of cool objects.  In this paper, we report on the result of mapping observations in molecular lines of $^{12}$CO, $^{13}$CO, C$^{18}$O, HCN, and CS, and in the 150 GHz continuum band using the Nobeyama 45-m telescope. The spatial and kinematic structures of the dust/molecular envelope around the central star have been revealed. In addition, we made photometric and imaging observations in the near-infrared $J$, $H$ and $K$ bands. We also made molecular-line observations by the other  lines, which clarify the special circumstellar chemistry of this object.  A part of the circumstellar chemistry of this object will, however, be presented in another paper.

%%%%% Observations and Results %%%%%
\section{Observations and Results}
\subsection{Observations}
Mapping observations in the CO $J=1$--$0$, $^{13}$CO $J=1$--$0$, C$^{18}$O $J=1$--$0$, CS $J=2$--$1$, and HCN $J=1$--$0$ lines were made with the 45-m telescope at Nobeyama during 2001 February 4--9. A cooled SIS-mixer receiver with a bandwidth of about 0.5 GHz was used, and the system temperature (including atmospheric noise) was 300 -- 500 K (SSB), depending on the frequency and weather. The aperture efficiencies at 86, 100, and 110 GHz were 0.41, 0.36, and 0.31, respectively. The main beam efficiencies at 86, 100, and 110 GHz were 0.49, 0.45, and 0.42, respectively. The half-power full widths of the main beam were 19$''$.8, 17$''$.1, and 15$''$.6 at 86, 100, and 110 GHz, respectively. The antenna temperature ($T_{\textrm{a}}$) given in the present paper was corrected for the atmospheric and telescope ohmic loss, but  not for the beam or aperture efficiency. Acousto--optical spectrometers with wide frequency coverage (AOS-W) were used. The AOS-W spectrometers have 250 MHz bandwidth and 2048 effective channels, the velocity coverage of about 800 km s$^{-1}$ and a spectral resolution of about 0.85 km s$^{-1}$ at 86 GHz. The rest frequencies of observing molecular lines were taken from \citet{lov92}.  For an intensity calibration  and a check of tuning frequencies, we observed molecular lines toward W51 (e.g., \cite{bel93}) prior to the observations of IRAS 19312+1950, and confirmed that the strong lines were surely detected toward this cloud. The telescope pointing was checked using a strong SiO maser source, V1111 Oph, and it was confirmed to be better than 3$^{''}$ under windless conditions. Mapping patterns for each line are summarized in table 1. The bipolar axis of this nebula seen on the 2MASS images is oriented approximately by 45$^{\circ}$ from the north (to the east). Therefore,  cross and rectangular  mappings were made along the perpendicular and parallel directions to the bipolar axis.  The IRAS position of this object (19$^{\textrm{h}}$ 31$^{\textrm{m}}$ 12$^{\textrm{s}}$.8, 19$^{\circ}$ 50$'$ 22$''$, B1950) was set as the map centers. Since contamination by background molecular emission was expected, we carefully chose the offset position for position switching; it was 20$'$ away from the mapping center (in the galactic latitude). The raw data were processed, flagging out bad scans, integrating with r.m.s.-weighted means, and removing the slopes in baselines.  

% t6-1.tex
{\setlength{\tabcolsep}{2pt}\footnotesize
%\begin{longtable}{llccl}
\begin{table}[hbtp]
\caption{Mapping patterns}
\begin{tabular}{llccl}
%\multicolumn{4}{c}{TABLE 1}\\
%\multicolumn{4}{c}{Mapping patterns}\\
\hline \hline
Molecule & Transition & Frequency   & Spacing & Pattern \\
                &                 &   (GHz) &    ($''$)   &    \\
\hline
%\endfirsthead
%\hline
%\endfoot
 % &  & ($^{''}$) &  \\
$^{12}$CO & $J=1$--$0$ & 115.27120 & 15 & 5$\times$5 \\ 
& & & 60 & 9 points cross \\
$^{13}$CO & $J=1$--$0$ & 110.20135 & 15 & 3$\times$3 \\
& & & 30 & 3$\times$3 \\
& & & 60 & 5 points cross \\
C$^{18}$O & $J=1$--$0$ & 109.78216 & 15 & 3$\times$3 \\
& &  & 30 & 9 points cross \\
HCN & $J=1$--$0$& 88.631847 & 15 & 3$\times$3 \\
& & & 60 & 5 points cross \\
CS & $J=2$--$1$& 97.98097 & 30 & 3$\times$3 \\
& & & 60 & 9 points cross \\
\hline
%\end{longtable}}
\end{tabular}
\end{table}
}

% table 2
%{\setlength{\tabcolsep}{3pt}\footnotesize
\begin{longtable}{clrrrrr}
%\multicolumn{7}{c}{TABLE 2}\\
%\multicolumn{7}{c}{Line parameters for IRAS 19312+1950}\\
\caption{Observed line intensities for IRAS 19312+1950}
\hline\hline
Molecule & Transition & $V_{\textrm{peak}}$ & $T_{\textrm{a1}}$ & r.m.s. & $T_{\textrm{a2}}$ & $I.I.$ \\
               &                 & (km s$^{-1}$) & (K) & (K) & (K) & (K km s$^{-1}$) \\
\hline
\endhead
\hline
\endfoot
\hline
\multicolumn{7}{l}{$T_{\textrm{a1}}$: peak intensity of original spectrum.}\\
\multicolumn{7}{l}{$T_{\textrm{a2}}$: peak intensity of background-subtracted spectrum.}\\
\multicolumn{7}{l}{$I.I.$: integrated intensity of background-subtracted spectrum.}\\
\endlastfoot
%$^{12}$CO & $J=1$--$0$ & 37.85 & 5.29 & 0.24 & 3.89 & 47.74 \\
$^{12}$CO & $J=1$--$0$ & 38.72 & 5.26  &  0.24  &  3.44  &  24.04 \\
%$^{13}$CO & $J=1$--$0$ & 36.22 & 3.73 & 0.05 & 3.35 & 29.26 \\
$^{13}$CO & $J=1$--$0$ & 36.22 & 3.73 & 0.05 & 3.35 & 19.24 \\
C$^{18}$O & $J=1$--$0$ & 35.60 & 1.05 & 0.06 & 0.90 & 3.78 \\
HCN & $J=1$--$0$ ($F=2$--$1$) & 36.03 & 0.60 & 0.04 & 0.50 & 2.26 \\
CS & $J=2$--$1$ & 35.50 & 1.11 & 0.05 & 0.92 & 5.96 \\
\end{longtable}
%}

The continuum observations at 150 GHz were made on 2001 January 11 and 12,  with the Nobeyama Bolometer Array (NOBA) mounted on the Nobeyama 45-m telescope. The NOBA is a seven-element bolometer array using a differential readout method (\cite{kun93}). The band pass of the filter was centered at 147 GHz with a bandwidth of 30 GHz. The half-power full width (HPFW) of the beam was 12$''$. Maps were made with raster scans along the right ascension and declination. The scanning speed was 15$''$ s$^{-1}$.  These maps were combined using the basket-weaving method to correct scanning effects. The angular size of the final map is $1'.9 \times 1'.9$ with a grid spacing of $5''.3$. The final integration time per sampling point was about 40 s. The r.m.s. noise level of the final map is $\sim$ 10 mJy beam$^{-1}$. Further details of the NOBA are given in \citet{kun93}. Pointing and flux calibrations were performed every hour by observing 3C 345 and NGC 7027. We assumed that the spectrum of 3C 345 is flat in the observing bandwidth. The effect of atmospheric extinction was corrected using the opacity of the atmosphere measured between observations. According to the uncertainty of these calibrations and the variation in the antenna efficiency due to wind, the uncertainty of the absolute flux density was estimated to be $\pm$30\%. 

Near-infrared photometric observations were made with the 2.3-m telescope of Australian National University at the Siding Spring Observatory, Australia, on 2000 July 7, using the infrared array camera CASPIR. The detector of the CASPIR was a 256$\times$256 InSb  array with the angular size of $0''.5$ per pixel at the sky, and it covers a field of view of 128$'' \times 128''$. The seeing size was 1$''$--2$''$. In order to prevent the saturation of the array, a short exposure time was used (0.3 s $\times$ 100 cycles for the $H$ and $K$ bands, 1 s $\times$ 100 cycles for the $J$ band, and 12 s $\times$ 10 cycles for the B$_{\gamma}$ and H$_2$ narrow bands). To determine the photometric magnitudes in the standard system, we took images of standard stars near IRAS 19312+1950. The data were reduced using the IRAF data-reduction package by a standard way: dark frames were subtracted from raw data; flat fielding was done with dome-flat frames and the magnitudes were measured by an aperture photometry technique using the DIGIPHOTO package in the IRAF. The raw images using the H$_2$ narrow-band filter (2.12 $\mu$m) exhibited a faint patchy structure a few second north of the central star. However, in the later analyses, this was turned out to be a ghost of the bright central star produced by reflection in the narrow-band filter. We detected no H$_2$ emission in this nebula.

\subsection{Results of Molecular Line Mappings}
Figure 1 shows the spectra of the CO $J=1$--$0$, $^{13}$CO $J=1$--$0$, C$^{18}$O $J=1$--$0$, HCN $J=1$--$0$, and CS $J=2$--$1$ transitions toward IRAS 19312+1950. The line parameters are given in table 2. We detected strong CO emissions toward IRAS 19312+1950. However, the 60$''$-grid 9-point cross mapping revealed that the emissions peaked at $V_{\rm lsr}=$27--33 and 43 km s$^{-1}$ are spatially extended more than a few arc minutes, indicating that these are from back/foreground molecular clouds. With the 15$''$ grid mapping, we found that the component at $V_{\rm lsr}=38$ km s$^{-1}$ is sharply peaked in intensity at the IRAS position.  The raw spectra indicate that  the spectral features of back/foreground clouds are relatively homogeneous over the mapping regions. Therefore, in order to obtain the spectra of the object,  we averaged the spectra at the 22$''$ to 60$''$ offset positions surrounding the object, and made an average background spectrum. We then subtracted it from the original on-source spectra and obtained the (supposedly) background-free spectra. The same procedures were applied to all of the lines, though contamination by background emission was not so severe except in the cases of the CO $J=1$--$0$ and $^{13}$CO $J=1$--$0$. The fore/background components of the CO $J=1$--$0$ line
were considerably flat and the intensity varies by only less than 10\% in the 1$'$ away spectra.
In the case of $^{13}$CO $J=1$--$0$, the background emission was not perfectly flat, so that the subtracted spectrum still exhibits emissions at $V_{\textrm{lsr}}=$28 and 45 km s$^{-1}$.  Figure 1 shows  the raw spectra at the mapping center (thin line), the average background spectra (broken line), and the background-subtracted spectra at the mapping center (bold line). In further analyses, we used the background-subtracted spectra to avoid the background contamination.

All of the spectra exhibit a clear narrow feature with a width of about 2--3 km s$^{-1}$ (hereafter narrow component) around $V_{\textrm{lsr}}= 36$ km s$^{-1}$. The widths of the narrow components seem to vary among the molecular lines. For instance, the lines of $^{12}$CO, $^{13}$CO, and C$^{18}$O have full widths at half maximum of about 3.0, 2.3, and 2.0 km s$^{-1}$, respectively.

On the other hand, the background-subtracted spectrum of $^{12}$CO (solid line in top left panel of figure 1) exhibits a weak broad feature with a line width of about 40 km s$^{-1}$.  This was quite weak and difficult to distinguish from noise or background emission.  However, in the top-left and right panels of figure 1,  we can clearly see emission at $V_{\rm lsr}=$ 20--25 and 48--55 km s$^{-1}$, which are significantly above the background emission (broken curve)\footnote{Though the presence of the broad component seems to remain somewhat debatable
in this single-dish observation, we have confirmed it with mm-wave interferometric observations.  The results described in this paper are completely consistent with the interferometric observations.}. The presence of this component can be seen more clearly in the HCO$^{+}$ spectrum, which is free from background contamination (\cite{deg03}). The broad component was also seen in the spectra of thermal lines of SO  and SiO (\cite{deg03}). In the case of HCN, we can also see a possible broad component, though it is not so clear because of the hyperfine splitting of the HCN $J=1$--$0$ transition: $F=0$--1, 2--1, and 1--1. The hyper fine lines should appear at $\Delta V= -7.44$ and $+4.68$ km s$^{-1}$ relative to the main $F=2$--$1$ line (89.08792 GHz) with an LTE (optically thin) line intensity ratios of 1:5:3 (e.g., \cite{cao93}). In the bottom left panel of figure 1, the origin of the radial velocity of the HCN spectrum is based on the rest frequency of the strongest transition $F=2$--1.

Figure 2 shows the integrated-intensity maps. In this figure, the $Y$-axis is taken along the bipolar axis of the nebula; it is inclined by 45$^{\circ}$ from the north. The background emission was subtracted in these maps. For making these maps, the data taken on different days (with different mapping patterns) were combined, because the wind often prevented us to continue mappings  (of course, the data taken under the wind speed more than 10 m s$^{-1}$ were discarded). In the cases of the $^{12}$CO and HCN lines, we made two separate maps: one for  the broad and another for the narrow components. We took the emission in the radial velocity ranges from 0 to 30 km s$^{-1}$ and from 50 to 60 km s$^{-1}$ for the broad component of the $^{12}$CO spectrum, and from 0 to 25 km s$^{-1}$, and from 45 to 60 km s$^{-1}$ for the broad component of the HCN. The velocity ranges of the broad component in the HCN spectrum were chosen by eye inspection. The other lines (except for $^{12}$CO and HCN) do not exhibit the broad feature; maps were made only for the narrow components. 

A remarkable point in figure 2 is that all of the intensity peaks of the narrow and broad components of molecules coincide with the IRAS position of IRAS 19312+1950. We do not find any differences in the peak position between the narrow and broad components, or between different molecular lines. The object seems to be slightly resolved by the 15$''$ beam. The directions of the elongation seem to be different between molecular lines; the narrow components of the $^{12}$CO and HCN spectra are somewhat elongated along the $X$-direction; the $^{13}$CO and C$^{18}$O spectra are elongate along the $Y$-direction; the broad component of the $^{12}$CO spectrum.

In order to evaluate the size of the source more carefully, we computed the model intensity variations and compared these with the observations. Figures 3a and 3b show  position vs. integrated-intensity plots along the X and Y directions for each line. The intensities expected  from simple homogeneous circular-disk models with radii of 0$''$ (point source), 10$''$, 20$''$, and 30$''$ are drawn in solid and broken curves alternatively. The horizontal and vertical tick marks indicate the uncertainty in telescope pointing and the rms of  integrated intensities, respectively.  In figure 3a, we can see that the extension of the narrow component of $^{12}$CO is clearly larger than that of the broad component. However, this phenomenon is not so clear in the case of HCN. It seems that the narrow components of all of the lines, except for that of HCN, extend to more than 15$''$ according to the simple homogeneous disk model. 
%n the cases of the $^{12}$CO and HCN spectra, we made the same plots for the broad and narrow components.

Figures 4a and 4b show position--velocity (PV) diagrams for each molecular line (background subtracted).  No particular spatial velocity gradient was seen in the $^{12}$CO and CS.   The narrow components ($\sim$ 35--36 km s$^{-1}$) of $^{13}$CO, C$^{18}$O, and HCN exhibit a small velocity gradient along the $X$-axis ($3.5 \times 10^{-2}$ km s$^{-1}$ arcsec$^{-1}$) that is perpendicular to the bipolar axis.  Even though the background emission was not completely homogeneous, the intensities of the background clouds at around 30 and 45 km s$^{-1}$ remained to be small compared with the narrow components, especially in the $^{12}$CO, $^{13}$CO, and C$^{18}$O lines. 

\subsection{Results of 150 GHz Continuum and Near-Infrared Observations}
Figure 5 shows the 150 GHz continuum map toward IRAS 19312+1950. 
The map was overlaid on the $JHK$  false-color composite image. 
The image size is $1'.75 \times 1'.75$. 
The 150 GHz continuum emission seems to extend to the north--east along the bipolar axis. The peak flux density of the 150 GHz continuum radiation was $7.7 \times 10^{-3}$ Jy per beam. 
The total integrated flux density of the source integrated over the inner 25$''$ was 0.07 Jy ($\pm 0.01$ Jy). The near-infrared image shows that the number density of bright stars is less at the northeast part of the object than at the southwest part, suggesting that the dust hides background stars more in the northeast part of the object. The magnitudes of the central star (excluding the ring-like elongated component) are 10.7, 7.8 and 6.3 mag (uncertainty: $\pm 0.1$ mag) in the $J$, $H$, and $K$-bands, respectively. 
Figure 6 shows the spectral energy distribution (hereafter SED) of IRAS 19312+1950. 
Black body curves are also drawn. The flux densities of 12, 25 60, and 100 $\mu$m were taken from the IRAS point source catalog.  This SED shows a flat (or moderate double peak) shape.  
An excess  can be seen between 60 and 100 $\mu$m. 
These two data points were fit by the blackbody curve of the temperature of 70 K. In the galactic plane, interstellar dust in background molecular clouds  occasionally emits far-infrared radiation. 
Therefore, this excess may possibly be caused by interstellar contamination. 
However, the IRAS mid- and far-infrared maps with about 1$^{'}$ resolution show a clear intensity peak at the IRAS position of IRAS 19312+1950. 
This fact suggests that most of the energy of radiation in the 60 and 100 $\mu$m bands originates from the central source of IRAS 19312+1950.

 %\twocolumn

%%%%% Discussion %%%%%
%%%%% Discussion %%%%%
\section{Discussion}
\subsection{Evidence of an Evolved Object}
Because molecular species  observed  toward this source are
somewhat similar to the molecules detected in dark clouds
(for example, in L 134N; \cite{swa92}), it is reasonable to doubt if this star
is a young stellar object with SiO masers as IRc 2 in Orion.
However, we think that this is not the case for IRAS 19312+1950.
Note that many molecular lines have also been detected in the protoplanetary nebula, OH 231.8+4.2
(\cite{mor87}).  Furthermore,  AGB maser stars have  occasionally been found in/toward 
molecular clouds (\cite{imai02})

Firstly, this object exhibits deep 2.3 $\mu$m CO first-overtone absorption bands 
corresponding to late M III (P. R. Wood 2002, private communication). 
Most of low-mass young stellar objects exhibit emission or weak absorption 
of the 2.3 $\mu$m CO overtone bands;  this fact is very often used to separate YSOs and AGB stars
(\cite{cas92}; \cite{cha93}). Exceptional cases are
IRc 2 in Orion and IRS 5 in L 1551, which exhibit the 2.3 $\mu$m CO absorption spectra 
somewhat similar to those of K I and K III, respectively \
(\cite{mor98}; \cite{car87}). However,  even in these exceptional YSOs,
the effective temperatures are much higher than those of M III stars.

Secondly, IRAS 19312+1950 is an isolated object.
We find no bright object within a few arcminutes from this star in the optical, near-infrared,
and middle-infrared (MSX) images. 
Moreover, we cannot find any star forming activities 
on the infrared images down to $K\sim 16$ mag
(note that interstellar extinction at $K$ band in this region is about 2 mag at most,
which is  estimated in subsection 3.2).
The closest IRAS source (IRAS 19309+1947) is located 5$'$ south-west.
We find several very faint MSX objects at about 4$'$ -- 10$'$ away
and radio continuum sources (probable H $II$ region)  about 10$'$
north-west (with the SIMBAD database). These objects are  located too far 
($\gtrsim$ 3 -- 7 pc at the distance of 2.5 kpc) to be considered as a cluster of 
star-forming activities associated with IRAS 19312+1950. They are probably fore/background objects.

Because both of above two reasonings are still not completely crucial
as evidence of an evolved star, we do not 
preclude the possibility of a young stellar object here. However,
because of the absorption feature of CO overtone bands,
the central star of IRAS 19312+1950 cannot be a low-luminosity young stellar object.
If we consider a high-luminosity ($\gtrsim 10^4 \; L_{\odot}$) young stellar object,
the distance and physical conditions must be somewhat similar to
those of AGB/post-AGB stars. Therefore, we postulate that
it is an AGB/post-AGB object in this paper, and estimate various quantities.
Further discussion on the possibility of a young stellar object will be given in a
future paper, which is based on the molecular species found in this object.

\subsection{Distance to the Object}
We can estimate the distance to the source from the SED of this object, 
applying an interstellar extinction law at each wavelength (for example, see \cite{mat90}).  
However,  the distance depends strongly on the amount of total extinction toward this particular object.
The extinction of the  CO fore/background cloud at $V_{\rm lsr}=27$ -- 33 km s$^{-1}$ is
estimated to be $A_V=16.3$ from the CO integrated intensity (24.7 km s$^{-1}$; 
for conversion factor,  see \cite{fre82}). If this value is used for the extinction assuming that 
 this is a foreground cloud, we obtain the distance of 2.5 kpc from the absolute luminosity 
 of the object, 8000 $L_{\odot}$, which is typical for AGB stars.
The bolometric correction to the IRAS 12 $\mu$m flux density (\cite{deg98}) was 
computed to be $(BC)_{12} = 3.8$, where halves of the IRAS flux densities at 60 and 100 $\mu$m 
are assumed to come from this object.
 
If we consider that the $V_{\rm lsr}=27$ -- 33 km s$^{-1}$ component cloud is behind the
object (though this is unlikely) and apply the standard interstellar extinction (for example, see \cite{deg98}),  
we obtain the distance of 3.9 kpc. The distance varies from 5.1 to 3.3 kpc, 
when all of the 60 and 100 $\mu$m flux densities come from the source, or nothing comes, respectively.  

The luminosity of the central star of PPNe is known to be larger than the assumed value, 8000 L$_{\odot}$. 
In such a case, the distance varies in proportion to  the square root of the luminosity.  On the other hand, 
the distance would be slightly smaller than the above value if we take into account the self-absorption 
due to dust in the narrow-component cloud.  Therefore,  we estimated that the distance ranges from 2.0  kpc to 5.1 kpc for this object.  If we compare this with  OH 231.8+4.2 (at the distance of 1.3 kpc; \cite{bow84}),   
IRAS 19312+1950 is still a bit a distant object, though the IRAS flux densities at middle infrared wavelengths are similar for both objects. 

For simplicity, we assume in this paper that the distance to the object is 2.5 kpc, and estimate various quantities.

\subsection{Broad Component in the $^{12}$CO Spectrum}
In the background-subtracted spectrum of $^{12}$CO (top two panels of figure 1), a broad component is seen in the velocity range $V_{\rm lsr} \sim $ 10 -- 60 km s$^{-1}$. This component is considered to be formed in the expanding envelope of the central mass-losing star. Because the broad component is weak and overlapped with the strong background emission, it is difficult to measure the accurate full line width of this weak component. The full width at zero maximum is estimated to be about 50 km s$^{-1}$, suggesting a typical expanding velocity of the envelope of 25 km s$^{-1}$. In addition, the velocity range of SiO maser emission, which is found in $V_{\rm lsr}=$20--55 km s$^{-1}$ (\cite{nak00}), is overlapped with the radial velocity of the CO broad component. Moreover,  the intensity map of the broad component (see top left panel of figure 2) shows a clear intensity peak at the center position of this object. These facts strongly suggest that the broad component is physically associated with the central star.  From the CO line intensity (and the intensities of other molecular lines), the mass loss rate can be calculated with the LVG model (for example, see \cite{mor77}). The best-fit value of the mass loss rate was calculated to be  $1.04\times 10^{-4} M_{\odot}$ yr$^{-1}$. The details of the LVG model calculation are to be given in a future paper.

\subsection{Narrow Components}
We should be careful in interpreting the narrow component. The width of the narrow component is about 3 km s$^{-1}$; this is nearly the same for all of the lines of $^{12}$CO, $^{13}$CO, HCN, and CS. On the other hand the widths of the narrow components of less abundant species (such as HCO$^{+}$, etc.) are dependent on molecular species (\cite{deg03}). This fact suggests that these molecular lines are formed in a turbulent molecular cloud (e.g., \cite{lad76}); the line width depends on the optical depth of the line in such a cloud. The narrow component emission may come from some cool component toward this object, for example, a dark cloud at fore/background of this source.

We can estimate a column density of H$_2$ from the CO line intensity when we assume that the narrow component of CO comes from a dark cloud.  Adopting the CO--H$_2$ conversion ratio of $2.6 \times10^{20}$ /(K km s$^{-1}$) (\cite{mag98}), we obtained a H$_2$ column density of $6.1 \times10^{21}$ cm$^{-2}$ from the present observation [$T_{\textrm{peak}}$($^{12}$CO)=3.9 K, HPFW=2.9 km s$^{-1}$, and the telescope beam efficiency of 0.48]. The total mass of H$_2$ within the 16$''$ telescope beam (assuming the distance of 2.5 kpc) is calculated to be 2.8 $M_{\odot}$. Alternatively, from the column-density estimate  using optically thin C$^{18}$O line (for example, see \cite{fre82}),  we obtain the mass of about 9.7 $M_{\odot}$ for the cloud within the 16$''$ telescope beam.  The mass estimates using CO lines are considered to be uncertain by a factor of few, depending on the isotopic abundance ratio. 
 
In the P--V diagram (figures 4a and 4b),  the narrow components of $^{13}$CO and C$^{18}$O exhibit a small velocity gradient along the direction perpendicular to the bipolar axis. We estimated the mass of the source based on this velocity gradient, assuming the Keplerian rotation of the narrow component,  0.5 km s$^{-1}$ at an inner radius of  $1.9 \times 10^{17}$ cm (5$''$). Then, we obtained about 3.5 M$_{\odot}$ as a mass of the object inside the radius of 7$''$. In fact, an inclination of the ring structure [which is seen with the SUBARU and UH88 telescopes (M. Tamura 2001, private communication)] is estimated to be $\sim67^{\circ}$. In this case the mass would increase by a factor of 1.2.

\subsection{Dust Emission}
We detected continuum emission at a level of 0.07 Jy at 150 GHz. This emission is considered to be radiation emitted by dust grains toward this object. 

We first estimate the mass of the cloud from a simple homogeneous-cloud model (e.g., \cite{hog97}). When we assume $T_{\textrm{dust}}=$ 10 K  and a distance of 2.5 kpc, we obtain the optical depth of the dust to be $1.4\times 10^{-3}$ at 150 GHz within the telescope beam. Adopting a mass-absorption coefficient ($\kappa_{\nu}$) of $3 \times 10^{-3}$ cm$^2$ g$^{-1}$ (including both gas and dust particles), we obtain 140 $M_{\odot}$ for the total mass of the dust cloud within the telescope beam. This value is inconsistent with the mass estimated from the CO narrow component, suggesting that the 150 GHz continuum emission does not come from the cold dark cloud. 

     Because the dust emissivity increases with the temperature of the grain, it is possible that the dust temperature is much higher than 10 K. In fact, the far infrared spectrum in figure 6 can be fit by a blackbody with $T_{\rm dust}$=70 K or higher. The dust temperature of an expanding envelope of a mass-losing star is higher near to the central star and decreases with the radius. \citet{kna93} calculated the continuum flux densities of stars with various mass-loss rates, and found that the model  flux densities fit well with the observed spectra. Their model for OH 231.8+1.4 with a mass-loss rate $2\times 10^{-4}\; M_{\odot}$ yr$^{-1}$ gave $F_{\textrm{150 GHz}}=0.39$ Jy at a distance of 1.3 kpc. If the same model is applied to IRAS 19312+1950, the flux density at 150 GHz is to be 0.1 Jy at the distance of 2.5 kpc. This value is roughly consistent with the continuum flux density that was obtained in the present paper. In this model, the total mass of the gas in the envelope within the radius of $9.3\times 10^{17}$ cm (25$''$) is computed  to be about  0.1 $M_{\odot}$ (corresponding to the broad component).

The dust is also responsible for the nebulosity seen on the near-infrared images, which is made by scattering the light from the central star.  We can estimate the amount of dust grains by assuming that the dust optical depth at the $K$ band is about unity in the envelope of this star.  In this case, a simple estimate of the mass of the the ring structure, which is present at the radius of 5$''$,   gives 0.9 $M_{\odot}$. Here, we assumed a 1$''$-thickness spherical dust shell and the standard gas-dust ratio. 

From these estimates of the masses in this and previous sections, and from the fact that the observed velocity of the narrow component ($\sim$35--37 km s$^{-1}$) coincides well with the center velocity of the broad components, we  conclude that the cloud responsible for the narrow component is not physically independent from the central star. Rather, we are compelled to consider that the narrow component cloud of a few solar mass is a part of the object. 

\subsection{Is IRAS 19312+1950 a Protoplanetary Nebula ?}
In general, PPNe are expected to have the following observational characteristics (e.g., \cite{kwo00}); (1) strong infrared excess that is due to circumstellar matter; (2) the effective temperature of the central star between F and G types, (3) the presence of a detached dust shell due to the termination of an AGB-phase mass loss, which gives a doubly peaked spectral energy distribution; (4) the nebulosity seen in scattered light and no line emission of highly ionized atoms. 

IRAS 19312+1950 shares the first and the forth properties. It exhibits a red IRAS color [$\log (F_{25}/F_{12})=0.47$] and shows a clear intensity peak at the IRAS position in the $^{12}$CO intensity map. The B$_{\gamma}$ narrow-band imaging with the ANU 2.3-m telescope  resulted no evidence of ionized gas. At mid-infrared wavelengths, no line emission feature was recognized in the IRAS low-resolution spectrum (except a probable noise at 17.8 $\mu$m; \cite{vol91}). On the other hand, we should be careful in characteristics (2) and (3). Effective temperatures of stars between AGB and PN phases are 5000 -- 7000 K, which correspond with the surface temperatures of F- or G-type stars.  In many cases, the central stars can be observable at optical wavelengths. A quite faint counterpart is recognizable at the position of the central star in the phase II DSS image in R-band, but the star image is too faint to judge the spectral type. One explanation of this discrepancy in the characteristics is that 
IRAS 19312+1950 suffers a large interstellar extinction of $A_V\sim$16 as estimated in subsection 3.2. 
%is at the stage just after the end of the AGB phase. In this case, the object should have almost the same effective temperature with that of AGB stars. The detection of SiO masers in IRAS 19312+1950 supports this explanation [e.g., \citet{lew89}]. 
The spectral energy distribution does not show clear single nor double peaks. However, if the interstellar extinction ($A_K=1.83$) is corrected, the fluxes at 0.9 -- 2.2 $\mu$m 
are increased by a factor of 30 -- 7, respectively. Therefore, the spectrum seems to be doubly peaked. The $J$-, $H$-, and $K$- band intensities are well fit by a blackbody with $T=1770$ K when the extinction of $A_K=1.83$ is corrected. The flux densities at 12 and 25 $\mu$m are well fit by a blackbody with $T=190$ K, which is a temperature reasonable for the dust shell at medium radii in the envelope (\cite{has94}). These properties also suggest that IRAS 19312+1950 is a very young protoplanetary nebula.
%Then, $R$-, $J$-, $H$- and $K$-band intensities are also well fit by the black body with the temperature of 1200K. 

Progenitors of O-rich protoplanetary nebulae are inferred to be relatively massive stars ($\geq $ 4 $M_{\odot}$).  According to the evolutionary scenario at the AGB phase, the stars with masses below $\sim$ 4 $M{_\odot}$ ends up as a C-rich star by the He-burning at the core and succeeding dredge-up. However, in stars above that, the created $^{12}$C atoms are further fused by the Hot Bottom Burning (HBB: \cite{ibe83}; \cite{woo83}) at the late-AGB phase  and the massive stars remain to be O-rich throughout the AGB phase with experiencing no (or a very short) C-rich era (\cite{mar01}). Therefore, the mass of the progenitor of IRAS 19312+1950 (O-rich)  is estimated to be relatively large. Recently, \citet{ven99} suggested that stars with a relatively large initial masses ($\sim$ 3.8 $M_{\odot}$) first go through a short ($\Delta t \sim$ 30000 yr) J-type ($^{13}$C-enhanced) carbon star era before the hot bottom burning  converts them back into oxygen-rich stars, and ends back again as C-rich stars. The estimated mass of the central star of IRAS 19312+1950 is somewhat too large to experience this evolutionary scenario.  

%However, it is possible that all of the OH/IR sources with very red colors ($C_{12}>0.5$) are not necessarily protoplanetary nebulae, but stars with a detached shell which is a result of short termination of mass loss after He-shell flash (Suh \& Jones 1997).  In this case, the termination of the mass loss of a certain period ($10^2$--$10^3$ yr) can explain the spectral energy distribution as shown in figure 5. In this case, a gradual recovery of the mass loss after the He-shell flash can produce SiO masers at the inner envelope. Therefore, the presence of SiO masers in IRAS 19312+1950 can be explained naturally. The defect of this hypothesis is that the mass loss rate must vary at least more than one order of magnitude compared with the rate during the normal period. The calculations made by Bl$\ddot{\rm o}$cker (1995) gave relatively small drop of the mass loss rate after the He-shell flash for high-mass stars ($> 5 M_{\odot}$). Therefore, it may be difficult to create the developed detached shell as found in IRAS 19312+1950.

The coincidence of the masses, about a few  $M_{\odot}$, which were calculated for the dust shell and the CO narrow component in the previous sections, even suggests us another idea that IRAS 19312+1950 is the AGB object embedded in a small globule-like cloud. The near-infrared radiation emitted by the central star can illuminate the dusty cloud having a mass of about a few $M_{\odot}$.  In this case, the cloud cannot be a cradle of the AGB progenitor because the mass of the progenitor is comparable with this cloud.  Only one possible case would be the AGB star colliding with the cloud (for example, a cometary cloud; see \cite{kul92}]. Even in such situation, the relative velocity between the cloud (the narrow component) and the AGB star (the broad component) is so small  ($\lesssim 1$ km s$^{-1}$) that it is unlikely for such event to occur often. 

%Strong evidence for CO
%Kleinmann & Hall sp atlas
%chandler 93
%cas92 CO
%bis97  confirmation 
%carr 1987 L1551 absorption KIII  is this real ?

%It has been suggested that the origin of bipolar nebulae is a binary system (Morris 1981), which is an attractive hypotheses to %explain the complexity of the envelope structure of IRAS 19312+1950. \citet{coh81} observed an excess of blue continuum toward %OH231.8+4.2, which is one of the secure precedents of protoplanetary nebula with SiO masers, suggesting the presence of a %companion to the Mira star. \citet{san00} noted that the bipolarity  and shocks of OH231.8+4.2 are indicators of the star that %recently entered the post-AGB phase, while the Mira variable is an AGB object. They proposed the possibility of a binary system %composed of an AGB and a post-AGB star. 

The line profiles of CO with broad and narrow features have been found in many stars (\cite{kna98}; \cite{ker99}).  These features are interpretated by multiple winds or Keplerian disk (\cite{ber00}). Therefore, it is possible to consider that the narrow-component cloud in IRAS 19312+1950 is  material trapped in an orbiting molecular reservoir (\cite{jur99}). It has been suggested 
that the origin of bipolar nebulae is a binary system [\cite{mor81}; \cite{coh81}; \cite{san00}].
However, in this case, it also  seems difficult to keep the ejected material of a few solar masses moving around the central star; precise modeling of this object is necessary. Because the trapped material (supposedly the ring structure) is located considerably far (5$''$) from the central star, the secondary star must be located also at the similar distance. However, we find no evidence for the other late-type star at such distances. Rather, it is possible that the counter part is evolved quickly to a white dwarf, and remnant of the ejected material still circles around the binary system, as suggested for the silicate carbon stars, etc. (\cite{llo90}; \cite{jur95}; \cite{kah98}). The existence of a hot secondary companion, however,  enables us to make models explaining the complex chemistry of IRAS 19312+1950.

With these arguments, we think at present that IRAS 19312+1950 is likely to be at an early stage of protoplanetary nebula. However, the origin of the narrow component is still a puzzle. The chemistry of this component is quite similar to that of dark clouds (\cite{deg03}), rather than that of protoplanetary nebula (\cite{mor87}), though the dust temperature seems too high ($\gtrsim$70 K). They may be a remnant of material, that is ejected in the AGB phase from the central star and is trapped by a distant binary counterpart.  Alternatively, a more drastic hypothesis may be a remnant of star forming material left after an encounter with a molecular cloud. More discussions on these points will be given in the future.

%%%%% Summary %%%%%
\section{Conclusion}
We made the mapping observations of IRAS 19312+1950 in molecular lines of CO $J=1$--$0$, $^{13}$CO $J=1$--$0$, C$^{18}$O $J=1$--$0$, CS $J=2$--$1$, and HCN $J=1$--$0$ lines and in the 150 GHz continuum. We also made photometric observations in the $J$, $H$, and $K$ bands to obtain the spectral energy distribution of IRAS 19312+1950. The main results are as follows:

\begin{itemize}
\item The line profile of $^{12}$CO (and possibly HCN) exhibits  weak broad ($\sim 40 $ km s$^{-1}$) and strong narrow ($\sim 3 $ km s$^{-1}$) components, whereas the profiles of the other lines ($^{13}$CO $J=1$--$0$, C$^{18}$O $J=1$--$0$, CS $J=2$--$1$) exhibit only narrow components. Both of the components show a clear intensity peak at the IRAS position of IRAS 19312+1950.

\item The narrow component spatially extends at least more than 15$^{''}$.

\item The spectral energy distribution of IRAS 19312+1950 exhibits a moderate double-peak profile (if extinction is corrected) at the near- and mid-infrared wavelengths with excess at 60 and 100 $\mu$m.
\end{itemize}

Based on the present observational results, we suggested that IRAS 19312+1950 is a  transient object between the AGB and planetary nebula phases, possibly a protoplanetary nebula. Though the present observations
do not completely exclude the possibility that this object is a young stellar object, evidence of an evolved object is accumulating furthermore with mm-wave interferometric observations.

\vspace{0.5mm}
The authors thank staff of Nobeyama radio observatory for the help of observations. They also acknowledge Peter Wood for the help in the near infrared observations at the Siding Spring Observatory and providing them the near-infrared spectrum of this object. This research is a part of the JN's thesis presented for the award of Ph.D. to the Graduate University for Advanced Studies, and is partly supported by Scientific Research Grant (C)(2) 12640243 of  Japan Society of Promotion of Sciences.

%\clearpage
%%% Figure 1 here
%\renewcommand{\thefigure}{fig1a}

% \twocolumn
\begin{figure*}
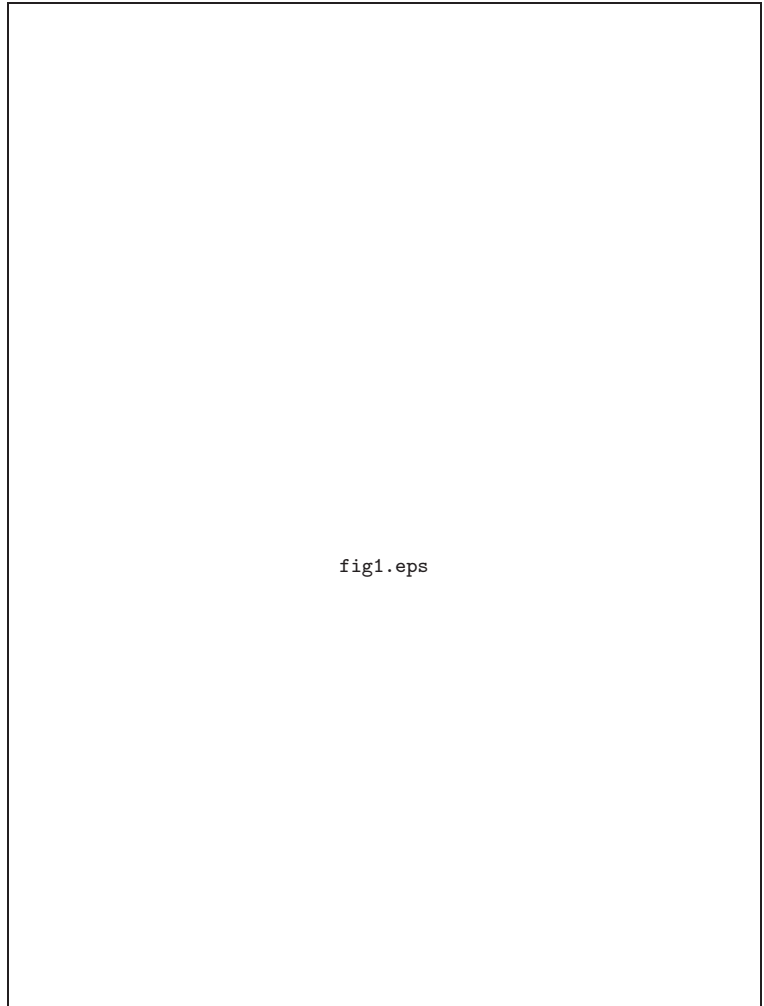

 \begin{center}
\FigureFile(100mm,150mm){fig1.eps}
 \end{center}
 \vspace{-2cm}
  \caption{Spectra of $^{12}$CO $J=1$--0 (top left),  enlargement of $^{12}$CO $J=1$--0 (top right),  $^{13}$CO $J=1$--0 (middle left), C$^{18}$O $J=1$--0 (middle  right), HCN $J=1$--$0$ (bottom left) and  CS $J=2$--$1$ (bottom right). The bold line indicates a background-subtracted spectrum, and the thin line indicates raw spectrum at the source center. In the panels of CO and $^{13}$CO, the broken line indicates the spectrum averaged over the spectra at $(X, Y) =$ (0$''$, $\pm$60$''$), ($\pm$30$''$, $\pm$30$''$), (0$''$, $\pm$30$''$), ($\pm$60$''$, 0$''$) and ($\pm$30$''$, 0$''$).  The spectra were averaged over at (0$''$, $\pm$60$''$), ($\pm$60$''$, 0$''$), ($\pm$30$''$, 0$''$) and (0$''$, $\pm$30$''$) for $^{18}$CO,  at  (0$''$, $\pm$60$''$), ($\pm$15$''$, $\pm$15$''$) and ($\pm$60$''$, 0$''$) for HCN, and at  (0$''$, $\pm$60$''$), ($\pm$30$''$, $\pm$30$''$), (0$''$, $\pm$30$''$), ($\pm$60$''$, 0$''$) and ($\pm$30$''$, 0$''$) for CS.  The broken line in the top right panel indicates the best fit parabola to the broad component.}
\end{figure*}

%%% Figure 2 here
\renewcommand{\thefigure}{2}
\begin{figure*}
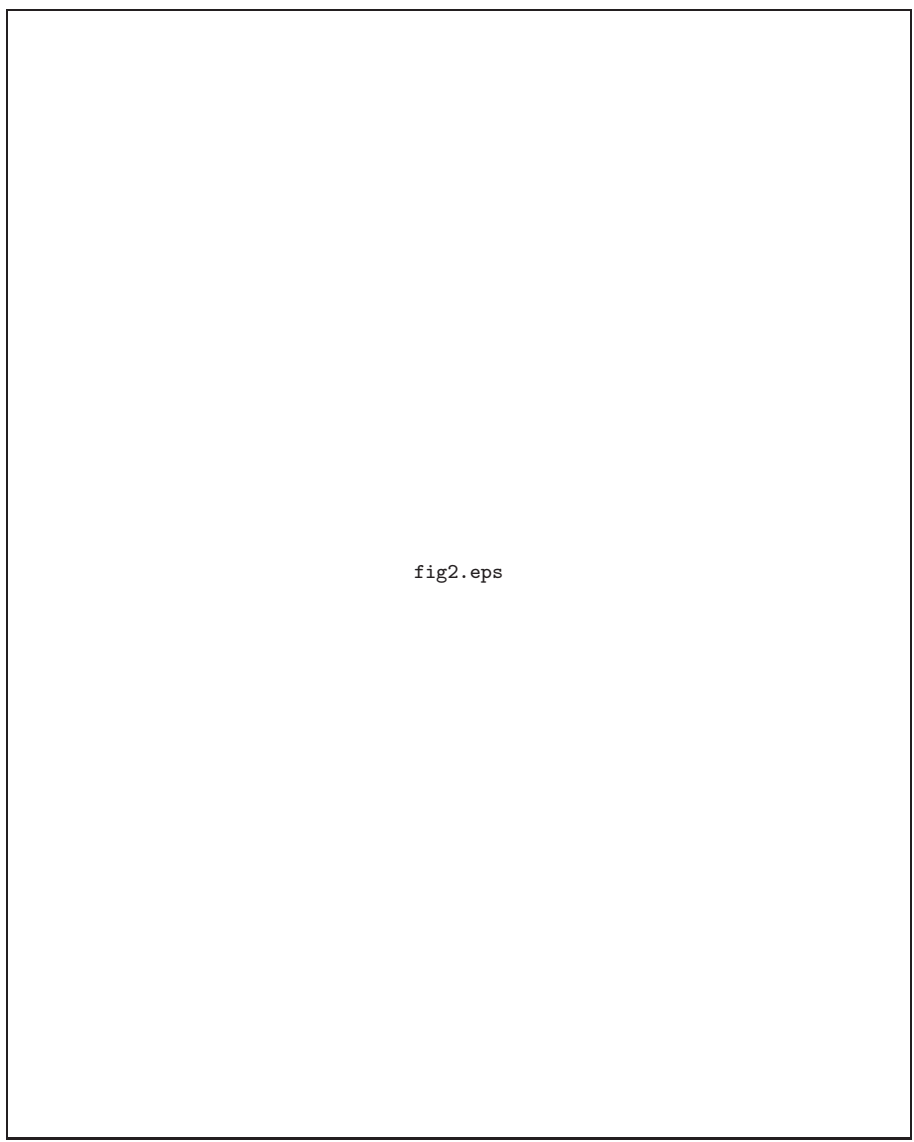

 \begin{center}
%%\FigureFile(60mm,60mm){fig2a-left.eps}
%\FigureFile(100mm,120mm){fig2.eps}
\FigureFile(120mm,150mm){fig2.eps}
 \end{center}
  \caption{Integrated intensity maps of the $^{12}$CO,  $^{13}$CO, C$^{18}$O, HCN and CS molecular lines. The broad (left) and narrow (right) components of $^{12}$CO and HCN are indicated by (B) and (N). Observed positions are indicated by the small dots.  The Y-axis is taken along the bipolar axis of the nebula, i.e., in the direction 45$^{\circ}$ from the north, and the X-axis that is perpendicular to it. The contour interval and the peak intensity are 1.0 and 14.3 K km s$^{-1}$, 1.0 and 18.3 K km s$^{-1}$, 2.0 and 29.2 K km s$^{-1}$, 0.5 and 5.7 K km s$^{-1}$, 0.5 and 5.9 K km s$^{-1}$, 0.1 and 3.2 K km s$^{-1}$ and 0.5 and 8.6 K km s$^{-1}$ for CO (B), CO (N), $^{13}$CO, C$^{18}$O, HCN (B), HCN (N) and CS, respectively. The beam sizes of the 45m telescope at 89 and 115 GHz are given in the bottom right panel.}
\end{figure*}

%%% Figure 3 here
\renewcommand{\thefigure}{3a}
\begin{figure*}
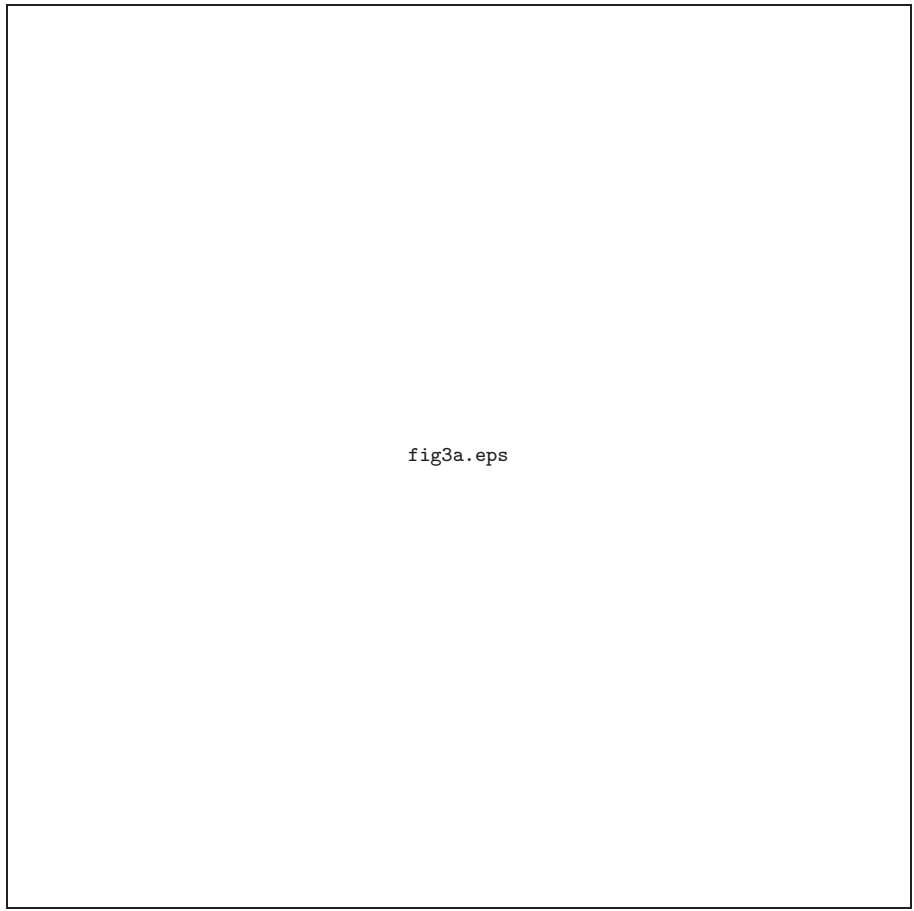

 \begin{center}
\FigureFile(120mm,120mm){fig3a.eps}
 \end{center}
  \caption{Spatial intensity variation of the broad and narrow components in the $^{12}$CO (left) and HCN (right) spectra. The filled and open circles indicate the data at offset positions along the X and Y directions, respectively. The solid and broken curves show the intensity variations that are expected from the simple circular disk model with radii of 0$''$ (point source), 10$''$, 20$''$ and 30$''$, alternatively. In the model, we assumed that the beam pattern of the telescope is Gaussian. The peak values of the model curves are adjusted to the observed values.}
\end{figure*}

\renewcommand{\thefigure}{3b}
\begin{figure*}
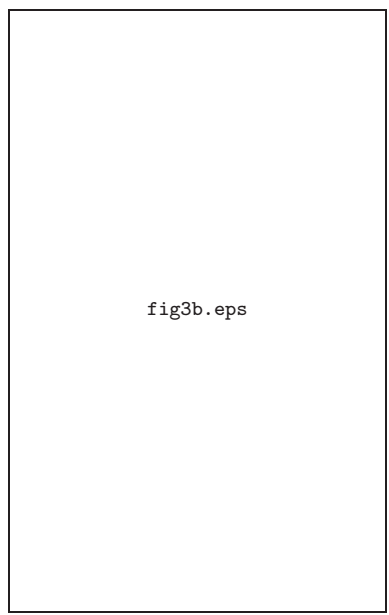

 \begin{center}
\FigureFile(50mm,80mm){fig3b.eps}
 \end{center}
  \caption{the same as figure 3a but the narrow components of $^{13}$CO (top),
 C$^{18}$O (center) and CS (bottom). }
\end{figure*}

%%%% Figure 4 here
\renewcommand{\thefigure}{4a}
\begin{figure*}
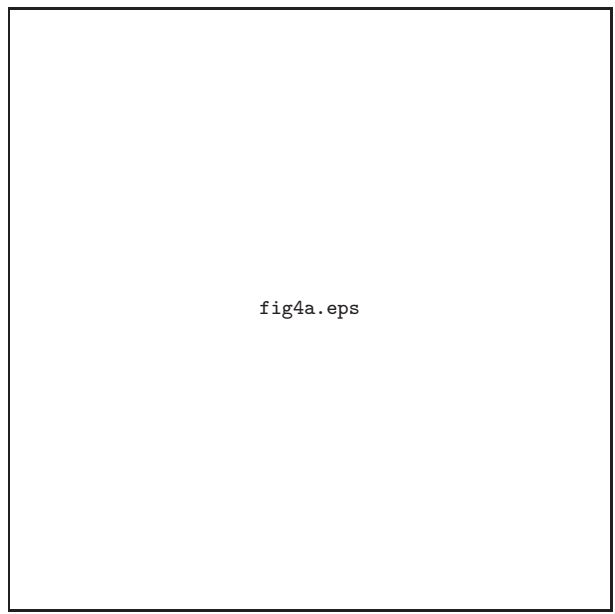

 \begin{center}
 \FigureFile(80mm,80mm){fig4a.eps}
 \end{center}
  \caption{Position--velocity diagrams for the $^{12}$CO (upper two panels), $^{13}$CO (Center two panels) and $^{18}$CO $J=1$--0 (lower two panels) spectra along the X (left) and Y (right) axes. The contour intervals and the peak intensities are 0.4 K and 3.4 K for $^{12}$CO, 0.5 K, and 3.4 K for $^{13}$CO and 0.1 K and 0.9 K for C$^{18}$O. }
\end{figure*}

\renewcommand{\thefigure}{4b}
\begin{figure*}
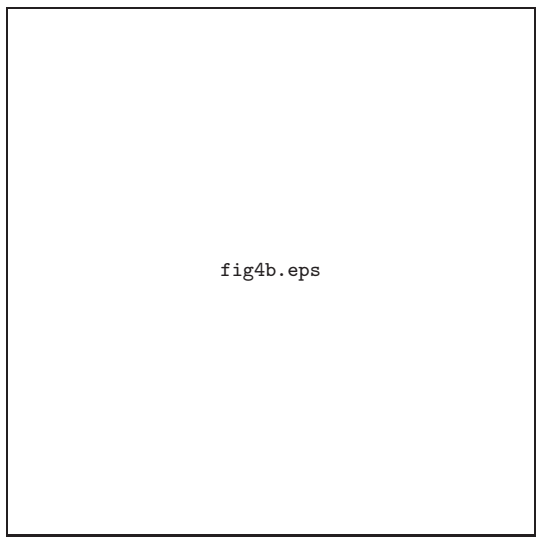

 \begin{center}
\FigureFile(70mm,70mm){fig4b.eps}
 \end{center}
  \caption{Same diagrams with figure 4a but for HCN (upper two panels) and CS (lower two panels). The contour intervals and the peak intensities are 0.05 K and 0.5 K for HCN and 0.1 K and 0.9 K for CS. }
\end{figure*}

%%%%%%%%%%%%%%%%%%%%%%%%%%%%%%%%%%%%%%%%%%%
\renewcommand{\thefigure}{5}
\begin{figure*}
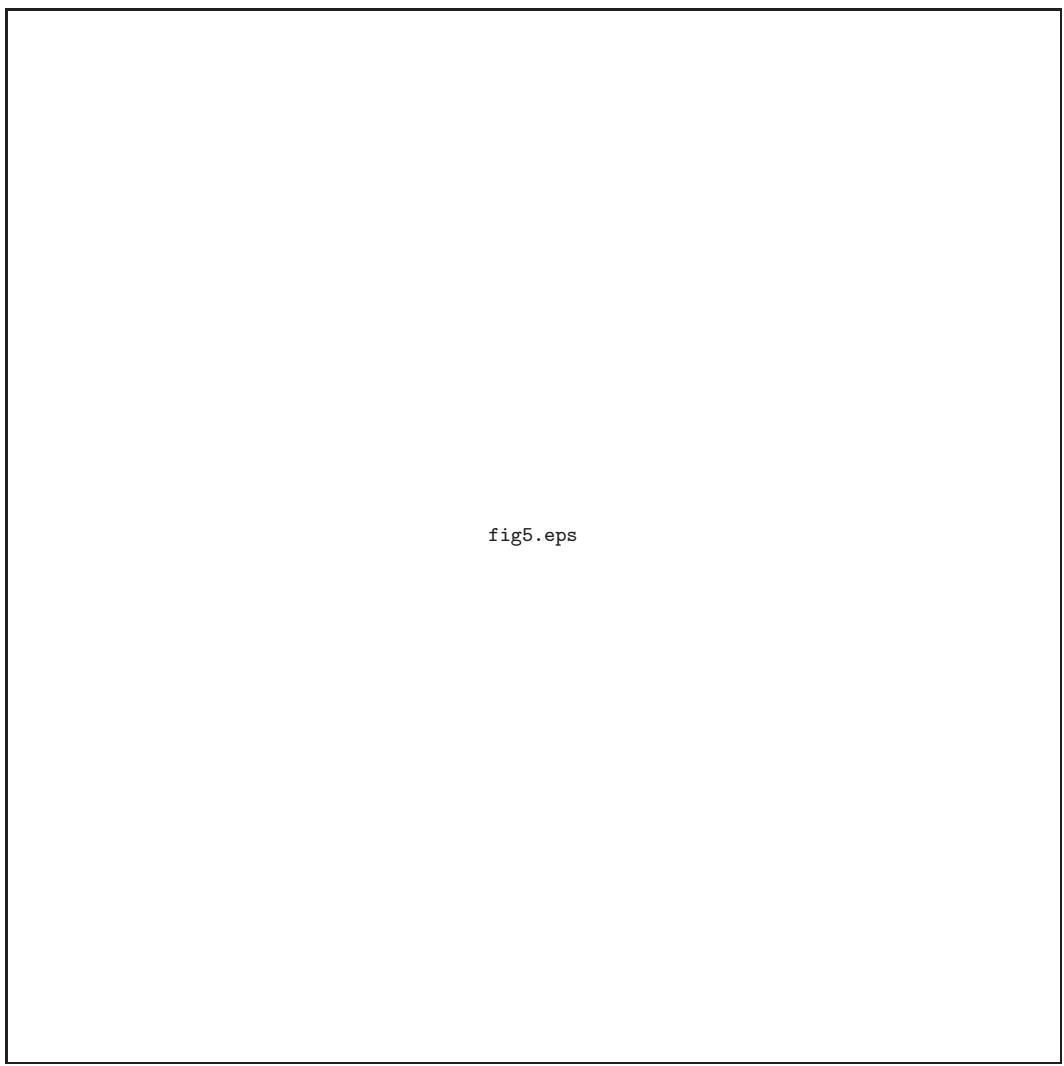

\begin{center}
\FigureFile(140mm,140mm){fig5.eps}
 \end{center}
 \caption{150 GHz continuum map (contour) toward IRAS 19312+1950 
 overlaid with the composite color image of the $J$, $H$ and $K$ bands (epoch: B1950).}
\end{figure*}

%%%% Figure 6 %%%%%
\renewcommand{\thefigure}{6}
\begin{figure}
 \begin{center}
\FigureFile(80mm,80mm){fig6.eps}
 \end{center}
  \caption{Spectral energy distribution of IRAS 19312+1950. The data at 0.65 $\mu$m (R-band) was taken from the DSS phase II archive. The circles between 12 and 100 $\mu$m indicate the flux densities from the IRAS point source catalog and the diamonds between 3.5 and 21 $\mu$m from the MSX catalog. The solid lines indicate expected lines from black body radiations at temperatures of 70, 190 and 1200K. The filled circles and filled squares indicate the cases of $A_V=3.6$ and 16.3 respectively.  In the case of $A_V=16.3$,  the extinction corrected flux density seems to be a bit high (open square).   
 }
\end{figure}

\end{document}